\documentclass[aps, pra, reprint, showpacs, twocolumn, 10pt, groupedaddress]{revtex4-1}

\usepackage{graphicx, amsmath, dsfont, dcolumn, bm, times, color, braket}

\begin{document}

\bibliographystyle{apsrev4-1}

\title{Phonon-induced decoherence of a charge quadrupole qubit}

\author{Viktoriia Kornich}
\affiliation{Department of Physics, University of Wisconsin-Madison, Madison, Wisconsin, 53706, USA}
\author{Maxim G. Vavilov}
\affiliation{Department of Physics, University of Wisconsin-Madison, Madison, Wisconsin, 53706, USA}
\author{Mark Friesen}
\affiliation{Department of Physics, University of Wisconsin-Madison, Madison, Wisconsin, 53706, USA}
\author{S. N. Coppersmith}
\affiliation{Department of Physics, University of Wisconsin-Madison, Madison, Wisconsin, 53706, USA}

\date{February  16, 2018}

\begin{abstract}
Many quantum dot qubits operate in regimes where the energy splittings between qubit states are large and phonons can be the dominant source of decoherence. The recently proposed charge quadrupole qubit, based on one electron in a triple quantum dot, employs a highly symmetric charge distribution to suppress the influence of charge noise. To study the effects of phonons on the charge quadrupole qubit, we consider Larmor and Ramsey pulse sequences to identify favorable operating parameters. We show that it is possible to implement typical gates with $>99.99\%$ fidelity in the presence of phonons and charge noise. \end{abstract}

\maketitle

\let\oldvec\vec
\renewcommand{\vec}[1]{\ensuremath{\boldsymbol{#1}}}

The past two decades have witnessed remarkable advances in the development of qubits in semiconductor-based systems \cite{petta:science05, bluhm:natphys11, kawakami:nnano14, veldhorst:nature15, goswami:natphys07, kawakami:pnas16, usman:nnano16, yoneda:arxiv17, loss:pra98, kane:nature98, divincenzo:nature00, khaetskii:prb00, hu:pra00, levy:prl02, taylor:natphys05}. Qubits constructed using electrons confined in lateral quantum dots (QDs) have been realized experimentally in different heterostructures for different types of qubits, including single-spin qubit \cite{kawakami:nnano14, veldhorst:nature15, kawakami:pnas16, yoneda:arxiv17, koppens:nature06, pioro-ladriere:natphys08}, singlet-triplet qubits \cite{dial:prl13, wu:pnas14}, quantum dot hybrid qubits \cite{shi:prl12, kim:nnano15, wang:arxiv17}, exchange-only qubits \cite{laird:prb10, medford:prl13, eng:science15}, and charge qubits with different numbers of electrons \cite{gorman:prl05, petersson:prl10, kim:nnano15, stockklauser:prl15}. The recently proposed charge quadrupole qubit, formed of one electron in a triple QD, is robust against uniform electric field fluctuations due to the high symmetry of its basis states \cite{friesen:ncom17, ghosh:prb17}. However, as for all qubits relying on a symmetric operating point, phonons can break this symmetry and cause decoherence. Indeed, phonons have been shown to be a significant source of decoherence for many types of QD qubits \cite{khaetskii:prb00, khaetskii:prb01, fedichkin:pra04, cheng:prb04, golovach:prl04, hu:prb05, stano:prl06, gamble:prb12, tahan:prb14, kornich:prb14, srinivasa:prb16}.

Here we study theoretically the decoherence of a charge quadrupole qubit arising from its coupling to phonons. In addition to decoherence between qubit states, we account for the effects of a leakage state. The leakage state is not coupled to the qubit subspace in the ideal case, however phonons break the symmetry of the qubit and consequently induce a coupling. The effects of phonons are greatest when the energy separation between the states is large, because of the rapidly growing  phonon density of states and the momentum-dependent electron-phonon matrix elements. To characterize the effects of phonon-induced decoherence and to identify the most favorable operating parameters for the qubit, we study both Larmor and Ramsey pulse sequences, which can be used to implement arbitrary single-qubit gate operations. The dependence of phonon-induced decoherence on qubit energy is found to be different from that of decoherence arising from charge noise \cite{ghosh:prb17, friesen:ncom17}, so both sources of decoherence must be considered to identify the optimal working regime for the qubit. By taking into account both of these processes, we identify a working regime where qubit fidelity can be greater than $99.99\%$.

\begin{figure}[tb] 
\begin{center}
\includegraphics[width=\linewidth]{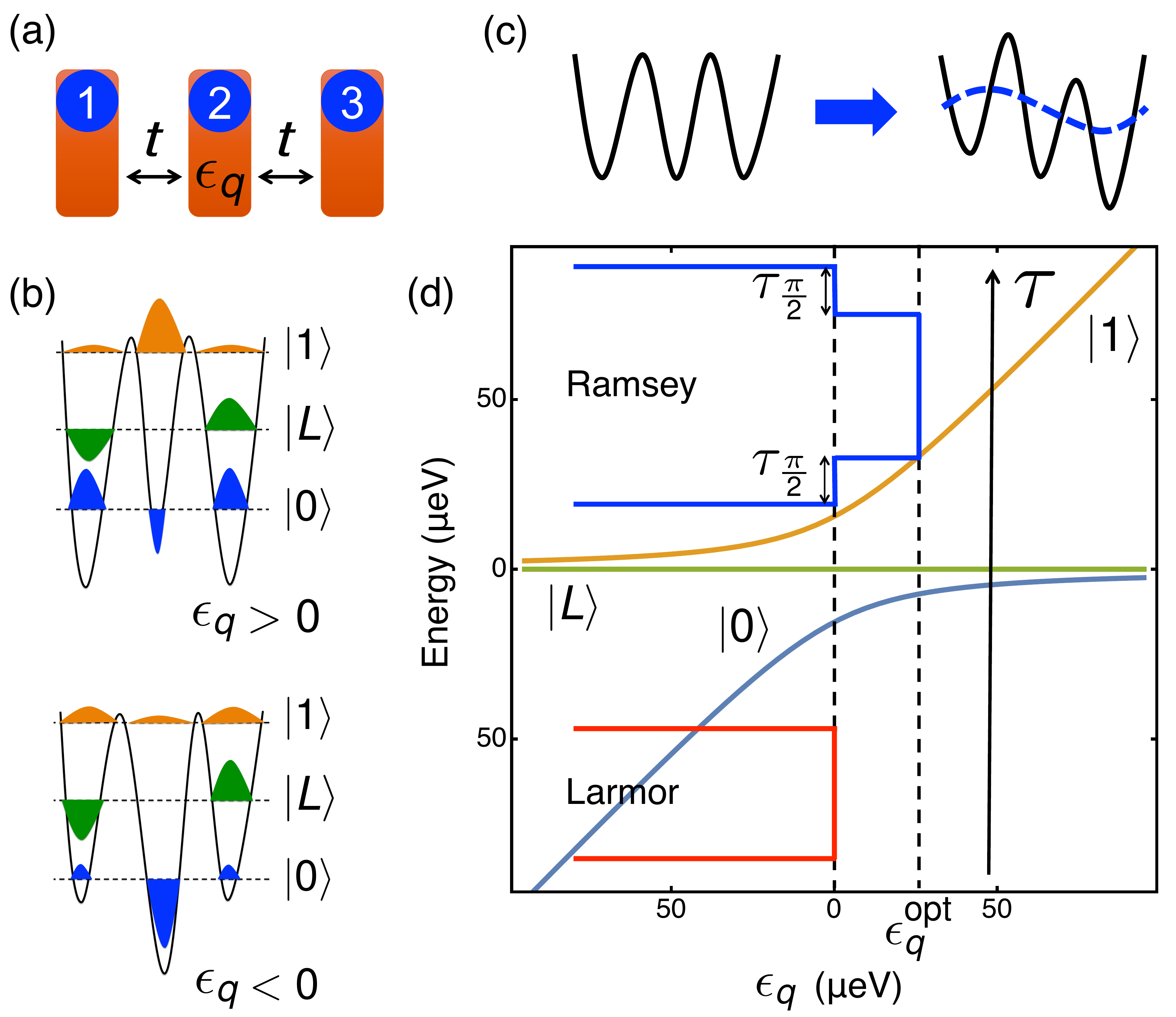}
\caption{(a) The charge quadrupole qubit consists of three quantum dots (blue circles), which are controlled by metallic top gates (red rectangles), and coupled via tunnel barriers. The quadrupolar detuning $\epsilon_q$ is controlled by the central gate. (b) A cartoon sketch of the wave functions of the basis states $|0\rangle$ (blue), $|L\rangle$ (green), and $|1\rangle$ (orange), for positive (top) and negative (bottom) $\epsilon_q$. (c) Qualitative illustration of the effect of a transverse phonon on the triple dot potential, resulting in decoherence. (d) The spectrum of a charge quadrupole qubit. Insets show the pulse sequences used to implement Larmor and Ramsey qubit operations studied here (vertical axis is time).}
\label{fig:pulses_setup}
\end{center}
\end{figure} 

The charge quadrupole qubit is formed in a triple QD that hosts one electron \cite{friesen:ncom17}, as illustrated in Figs.~\ref{fig:pulses_setup} (a) and \ref{fig:pulses_setup} (b). The qubit is operated in a highly symmetric fashion, such that the energies of the first and third QDs are the same and the middle QD has the same tunnel coupling to the two outer QDs. However, phonons can perturb the potential of the triple QD, as shown in Fig.~\ref{fig:pulses_setup} (c), giving rise to decoherence. To model this system, we consider electron wave functions in the $x-y$ plane of a quantum well, which are constructed from the ground states of harmonic oscillator potentials that approximate the QD confinement \cite{burkard:prb99}. For the left and right QDs these are given by $\psi_{L,R}=1/(\sqrt{\pi}l)\exp[-((x\pm a)^2+y^2)/(2l^2)]$, while $\psi_C=1/(\sqrt{\pi}l)\exp[-(x^2+y^2)/(2l^2)]$ for the central dot, where $a$ is the interdot distance and $l$ characterizes the radius of the electron wave functions. Here, $a$ and $l$ are assumed to be the same for all three QDs. To take into account the tunneling between the dots, we calculate the overlaps $S=\langle\psi_L|\psi_C\rangle=\langle\psi_R|\psi_C\rangle$. Here we may neglect the overlap between $\psi_L$ and $\psi_R$, as they are far apart. We now construct the orthonormal wave functions for electrons in the left or right QDs \cite{mattis:book81, burkard:prb99}, $\Phi_{L,R}=(\psi_{L,R}-2S\psi_C)\varphi$, and the central dot $\Phi_C=(\psi_C-g\psi_L-g\psi_R)\varphi/\sqrt{1-4gS+2g^2}$, where $g=S/(4S^2-1)$, and $\varphi(z)$ represents the separable component of the wave function in the heterostructure growth direction \cite{suppl}. Below, we specifically consider the case of Si heterostructures, which possess both orbitally excited states and excited states of the conduction band valleys \cite{zwanenburg:prm13, goswami:natphys07, xiao:apl10, yang:ncom13}. In our quantum dot basis set, we assume that such excited states are well separated energetically. However, in some situations, these states could cause undesired leakage channels.

The Hamiltonian of an electron in a triple QD coupled to the bath of phonons is given by $H=H_0+H_{el-ph}+H_{ph}$, where $H_0$ is the Hamiltonian of an electron in a triple QD, $H_{el-ph}$ describes the electron-phonon interaction, and $H_{ph}$ is the phonon bath Hamiltonian. Evaluating $H$ in the $\{|\Phi_L\rangle, |\Phi_C\rangle, |\Phi_R\rangle \}$ basis yields
\begin{equation}
\label{eq:Hamiltonian}
 H=\begin{pmatrix}\epsilon_d && t_A && 0\\
 t_A && \epsilon_q && t_B\\
 0 && t_B && -\epsilon_d\end{pmatrix}+\begin{pmatrix}P_{LL} && P_{LC}&& 0\\
 P_{LC}^\dagger && P_{CC} && P_{CR}\\
 0 && P_{CR}^\dagger && P_{RR}\end{pmatrix}+H_{ph},
\end{equation}
where $t_A$, $t_B$ are tunnel couplings between the dots, $\epsilon_q$ is the quadrupolar detuning \cite{friesen:ncom17}, $\epsilon_d$ is the dipolar detuning, and $P_{LL}$, $P_{CC}$, $P_{RR}$, $P_{LC}$, $P_{CR}$ are electron-phonon interaction matrix elements, defined as $P_{\alpha,\beta}=\langle\Phi_\alpha|H_{el-ph}|\Phi_\beta\rangle$. Note that we neglect the matrix elements $P_{LR}$ here because the corresponding wave functions have a negligible overlap. When  $t_A=t_B=t$ and $\epsilon_d=0$,  $H_0$ [the first term in Eq. (\ref{eq:Hamiltonian})] describes a charge quadrupole qubit \cite{friesen:ncom17}, whose eigenstates form a three-dimensional space in which two states $|0\rangle$ and $|1\rangle$ serve as the qubit basis, and the remaining leakage state $|L\rangle$ is decoupled from the qubit subspace in the absence of environmental noise. See the schematic illustration of the wave functions for negative and positive $\epsilon_q$ in Fig.~\ref{fig:pulses_setup} (b). The electron-phonon interaction is \cite{yu:book10, herring:pr56, kornich:arxiv16} $H_{el-ph}=\sum_{{\bm q},s}W_{s}(\bm{q})b_{{\bm q},s}e^{i{\bm q}\cdot{\bm r}}+\mbox{h.c.}$, where ${\bm r}$ denotes the position of the electron, ${\bm q}$ is the phonon wave vector, $s$ labels the longitudinal and two transverse modes, and $b_{{\bm q},s}$ is the phonon annihilation operator. The function $W_s({\bm q})$ depends on the bulk mass density, the speed of sound for mode $s$, and the deformation potential constants \cite{suppl}. In this work, we assume the following parameters for Si: bulk deformation potential constants $5$ eV and $8.77$ eV \cite{yu:book10}, speed of sound for longitudinal phonons $v_l=9\times 10^3$m/s and transverse phonons $v_{t1}=v_{t2}=5.4\times 10^3$ m/s \cite{cleland:book03, adachi:book05}, the bulk mass density $2.33$ g/cm$^3$.   
The full width at half maximum of $\varphi(z)$ is $4.7$ nm.

\begin{figure}[tb] 
\begin{center}
\includegraphics[width=0.828\linewidth]{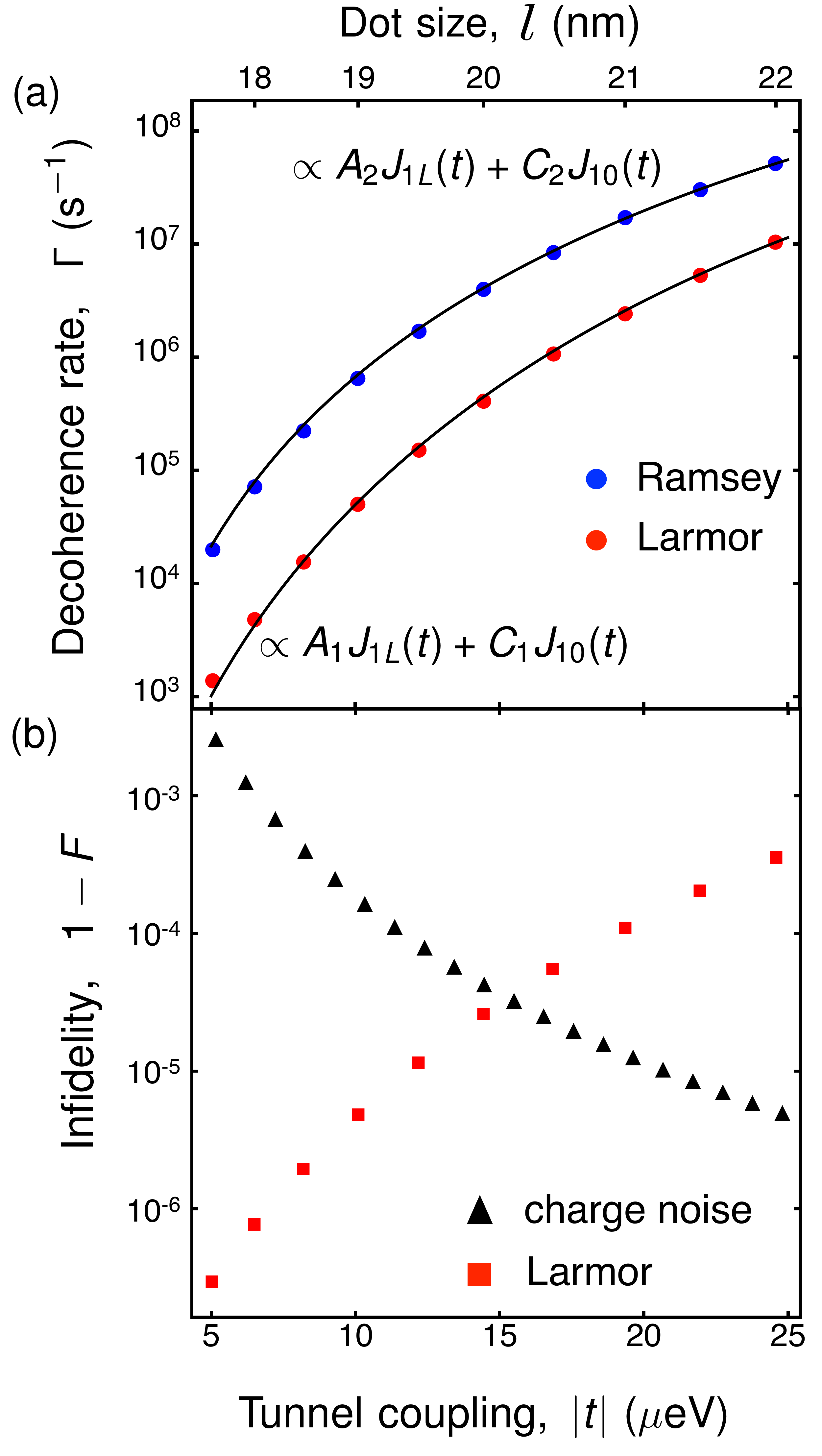}
\caption{(a) Dependence of the phonon-induced decoherence rates on the tunnel coupling magnitude $|t|$ for the Larmor  (red dots) and Ramsey (blue dots) sequences shown in Fig.~\ref{fig:pulses_setup} (d). Here, the interdot separation is $a=90$ nm, the dot size $l$ (top axis) is determined transcendentally from Eq. (\ref{eq:TunnelCoupling}), and $T=50$ mK. The dominant decoherence processes involve transitions between $|1\rangle$ and $|L\rangle$ and between $|1\rangle$ and $|0\rangle$, with approximate behaviors at $\epsilon_q=0$ given by the functions $J_{1L}(t)$ and $J_{10}(t)$ as defined in the main text. To demonstrate the usefulness of these functions, we fit the data using the forms shown in (a), yielding the following fitting coefficients: $A_1=50.95$, $C_1=102$ and $A_2=1338.12$, and $C_2=50.29$.  (b) Average gate infidelity $1-F$ of the free evolution portion of the Larmor pulse sequence (red squares), computed in the presence of phonons. The infidelity grows with $|t|$, so high-fidelity operations can be achieved only when $|t|$ is not too large. The charge noise-induced infidelity (black triangles) is computed, as described in Refs. \cite{ghosh:prb17, suppl}, and shows the opposite trend with $|t|$. Taking both noise processes into account, an optimal working point emerges at $|t|=15\ \mu$eV.}
\label{fig:L_t_dependence_BR}
\end{center}
\end{figure}

In addition to the QD energies appearing in $H_0$, several other experimental parameters can be tuned in an ideal device, including the geometrical parameters $a$ and $l$, and the temperature $T$. Changing $a$ or $l$ can affect the tunnel coupling $t$ between the dots. To estimate the value of $t$, we model $t$ as tunnel coupling between the two sides of a double well potential $\hbar^2(x^2-a^2/4)^2/(2m_{\rm eff}l^4a^2)$, where $m_{\rm eff}=1.73\times 10^{-31}$ kg is the transverse effective mass of an electron in Si, yielding \cite{suppl}
\begin{equation}
\label{eq:TunnelCoupling}
t=-\frac{3\hbar^2(a^2+4l^2)}{64m_{\rm eff}l^4\sinh{[a^2/(4l^2)]}},
\end{equation}
from which we see that $|t|$ grows rapidly with $l$.  

To investigate the time evolution and decoherence of the qubit, we employ the Bloch-Redfield formalism \cite{blum:book96, xu:pra14, golovach:prl04}. First, we apply the transformation $U_d$ that diagonalizes $H_0$ to the full Hamiltonian $H$, yielding $\tilde{H} =U_d^{-1}HU_d$. Note that the resulting Hamiltonian is expressed in the $\{|0\rangle,\ |1\rangle,\ |L\rangle\}$ basis. Similarly, we define $\tilde{H}_{el-ph} =U_d^{-1}H_{el-ph}U_d$. We then derive the equation for the time evolution of the  density matrix of our three-level system following the procedure described in Ref. \cite{blum:book96} (see Ref. \cite{suppl}): (1) write down the von Neumann equation, (2) trace over phonon degrees of freedom, (3) apply the Born-Markov approximation. This yields the following equations of motion for the elements of the density matrix $\rho$: 
\begin{eqnarray}
\label{eq:BlochEquation}
\nonumber
\dot{\rho}_{nm}=-i\omega_{nm}\rho_{nm}+\sum_{k,j}[(\Gamma_{jmnk}^++\Gamma_{jmnk}^-)\rho_{kj}\\-\Gamma_{kjjm}^-\rho_{nk}-\Gamma_{nkkj}^+\rho_{jm}],\end{eqnarray}
where 
\begin{subequations}
\begin{equation}
\label{eq:Gammaplus}
\Gamma_{jmnk}^+=\frac{1}{\hbar^2}\int_0^\infty d\tau \langle\bar{H}_{el-ph}^{jm}(\tau)\bar{H}_{el-ph}^{nk}(0)\rangle e^{-i\omega_{nk}\tau}, 
\end{equation}
\begin{equation}
\label{eq:Gammaminus}
\Gamma_{jmnk}^-=\frac{1}{\hbar^2}\int_0^\infty d\tau \langle\bar{H}_{el-ph}^{jm}(-\tau)\bar{H}_{el-ph}^{nk}(0)\rangle e^{-i\omega_{jm}\tau}. 
\end{equation}
\end{subequations}
Here, $\Gamma^+_{jmnk}$ and $\Gamma^-_{jmnk}$ are decoherence rates due to the electron-phonon interaction, the indices $n,m,j,k$ refer to the eigenstates of $H_0$ $\{|0\rangle,\ |1\rangle,\ |L\rangle\}$, the frequency $\omega_{nm}$ is the splitting between states $n$ and $m$, and $\tau$ is time. The angular brackets denote an average over phonon degrees of freedom, $\langle \hat{O}\rangle=\mbox{Tr}_{ph}\{\hat{O}\rho_{ph}\}$, where $\rho_{ph}$ is the thermal equilibrium state of the phonon bath. The notation $\bar{H}_{el-ph}(\tau)$ indicates that $\tilde{H}_{el-ph}$ is expressed in the interaction representation, defined as $e^{iH_{ph}\tau}\tilde{H}_{el-ph}e^{-iH_{ph}\tau}$.

After calculating the rates $\Gamma^-_{jmnk}$ and $\Gamma^+_{jmnk}$, we solve Eq. (\ref{eq:BlochEquation}) for the Larmor and Ramsey pulse sequences shown in Fig.~\ref{fig:pulses_setup} (d). For both sequences, we take the initial state to be the eigenstate $|0\rangle$, defined at $\epsilon_q=-0.08$ meV. For the Larmor pulse sequence the qubit is suddenly pulsed to $\epsilon_q=0$, where it evolves for time $\tau$, and is then pulsed back to $-0.08$ meV, where it is measured. For the Ramsey pulse sequence, we first pulse to $\epsilon_q=0$ and perform an $X(\pi/2)$ rotation. Since decoherence during Larmor rotations has already been addressed, we do not include it in this part of the Ramsey evolution. The system is then pulsed to $\epsilon_q=\epsilon_q^{\rm opt}$, corresponding to a rotation about the $(\hat{z}+\hat{x})/\sqrt{2}$ axis on the Bloch sphere. There are two reasons for choosing this value of $\epsilon^{\rm opt}$. First, a perfect $Z$ rotation requires pulsing $\epsilon_q\rightarrow\infty$, which is not experimentally feasible; a more realistic approach is to perform a three-step sequence that yields an effective $Z$ rotation \cite{hanson:prl07}. Second, as we show below, larger values of $\epsilon^{\rm opt}$ yield faster decoherence, so we choose the minimum value of $\epsilon^{\rm opt}$ that is consistent with \cite{hanson:prl07}. The system evolves freely at $\epsilon_q=\epsilon^{\rm opt}$ for time $\tau$, where it experiences phonons. We then pulse back to $\epsilon_q=0$ for another $\pi/2$ rotation and back to the base detuning $\epsilon_q=-0.08$ meV. We define decoherence rates $\Gamma_{\rm Lar}$ and $\Gamma_{\rm Ram}$ for the Larmor and Ramsey pulse sequences, respectively by fitting the decay of the resulting density matrix elements as a function of evolution time to the exponential form $e^{-\Gamma\tau}$.    

\begin{figure}[tb] 
\begin{center}
\includegraphics[width=\linewidth]{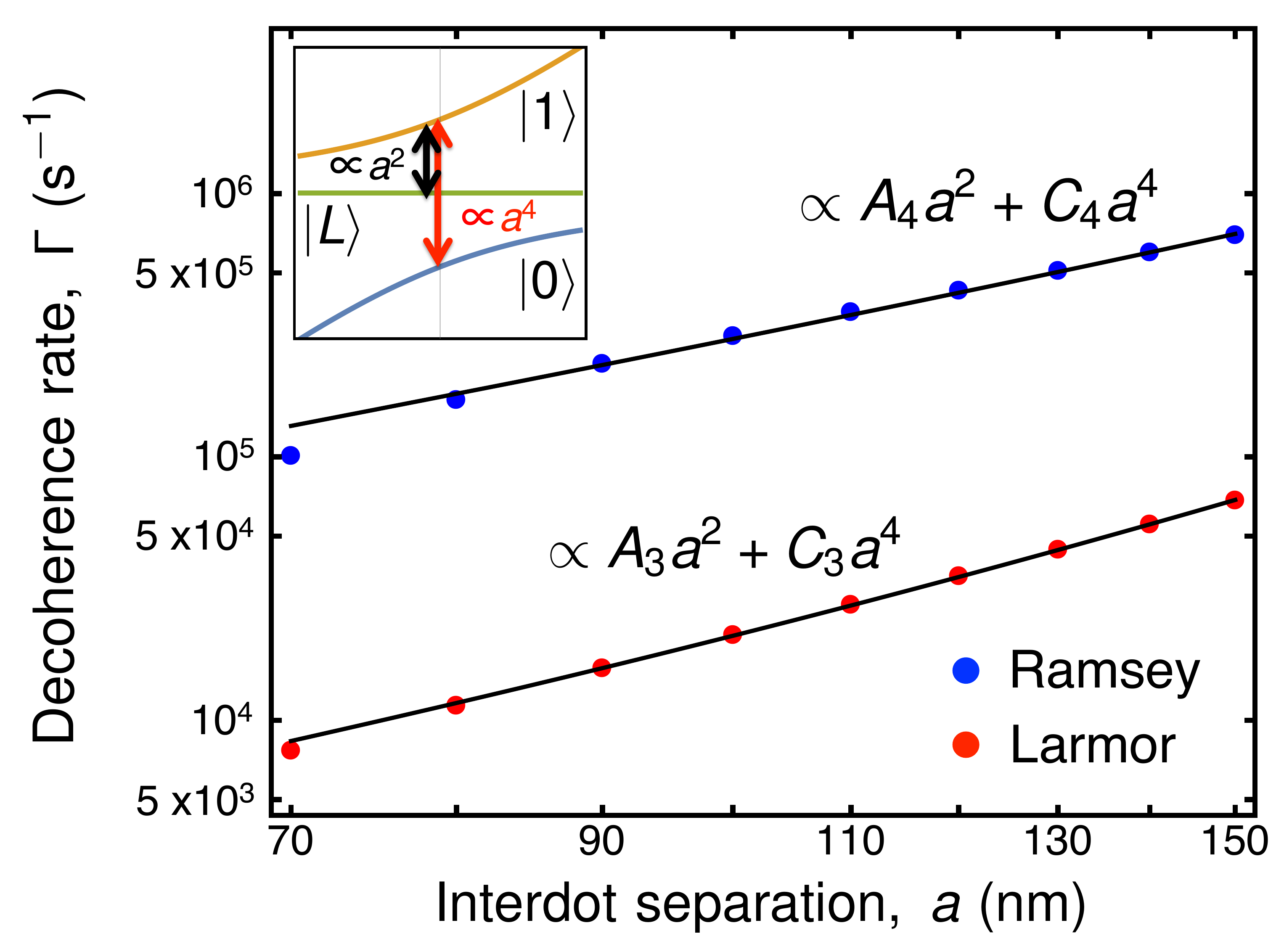}
\caption{Dependence of the phonon-induced decoherence rates on interdot separation for the Larmor (red dots) and Ramsey (blue dots) pulse sequences, keeping the tunnel coupling fixed at $t=-8.2\ \mu$eV. The decoherence rates grow with $a$ because the same long wavelength phonon causes a larger distortion of the triple dot. The specific dependences of decoherence rates on $a$, for the dominant $|1\rangle\leftrightarrow|L\rangle$ and $|1\rangle\leftrightarrow|0\rangle$ decoherence processes shown in the inset, are given by $a^2$ and $a^4$ respectively. We therefore fit the numerical results to the form shown in the figure, yielding the fitting coefficients $A_3=1.33$, $C_3=0.000077$, $A_4=25.52$, and $C_4=0.000255$. Here, $l=18.5$ nm and $T=50$ mK.}
\label{fig:interdot_dist_dependence_BR}
\end{center}
\end{figure}

Figure \ref{fig:L_t_dependence_BR} (a) shows the dependence of $\Gamma_{\rm Lar}$ and $\Gamma_{\rm Ram}$ on $l$ and $|t|$. Here, the growth of the rates is mainly due to the growth of $|t|$; the change in $l$ apart from its effect on $|t|$ is insignificant. The dominant decoherence processes for both pulse sequences are the transitions between $|1\rangle$ and $|0\rangle$, and between $|1\rangle$ and $|L\rangle$, as indicated in the inset of Fig.~\ref{fig:interdot_dist_dependence_BR}. The main contribution comes from the parts of the corresponding matrix elements of $\tilde{H}_{el-ph}$ containing terms $P_{RR}-P_{LL}$ for $|1\rangle\leftrightarrow|L\rangle$ and $P_{LL}+P_{RR}-2P_{CC}$ for $|1\rangle\leftrightarrow|0\rangle$. Consequently, for $\epsilon_q=0$ (i.e., Larmor oscillations) the decoherence rate is proportional to the function $J_{1L}(t)=|t|^3|\exp[\frac{-ia\sqrt{2}|t|}{\hbar v_{t1}}]-\exp[\frac{ia\sqrt{2}|t|}{\hbar v_{t1}}]|^2$ for the $|1\rangle\leftrightarrow|L\rangle$ process or $J_{10}(t)=|t|^3|\exp[\frac{-ia2\sqrt{2}|t|}{\hbar v_{t1}}]+\exp[\frac{ia2\sqrt{2}|t|}{\hbar v_{t1}}]-2|^2$ for the $|1\rangle\leftrightarrow|0\rangle$ process. (For details, see \cite{suppl}.) Here, $v_{t1}$ is the speed of sound for transverse acoustic phonons, that dominate over longitudinal phonons for the parameters used. The origins of $J_{10}(t)$ and $J_{1L}(t)$ are easy to understand: $|t|^2$ corresponds to the phonon density of states, another $|t|$ comes from the form of the deformation potential, while the exponential terms reflect the phase differences between the different matrix elements, which arise from the dot separations. Figure \ref{fig:L_t_dependence_BR} (a) shows that the calculated $\Gamma_{\rm Lar}$ is well described by the function $A_1J_{1L}(t)+C_1J_{10}(t)$, with fitting constants $A_1$ and $C_1$. The dependence of $\Gamma_{\rm Ram}$ also fits well to the same functional form, as shown in Fig.~\ref{fig:L_t_dependence_BR} (a), even though its dependence on $|t|$ is more complicated. 

One of the main questions to be answered in the present work is whether phonon-mediated noise presents a challenge for charge quadrupole qubits, and how this compares to charge noise, which was studied in Refs. \cite{friesen:ncom17, ghosh:prb17}. Figure \ref{fig:L_t_dependence_BR} (b) shows our main results for both types of noise.  Here the charge noise was assumed to be quasistatic, with a typical experimental noise value $1\ \mu$eV \cite{thorgrimsson:qi17}, affecting the dipolar detuning parameter $\epsilon_d$ as described in \cite{ghosh:prb17}. A composite $Z(\pi)X(3\pi)Z(-\pi)$ pulse sequence was implemented, to eliminate the effects of leakage to leading order and the average gate fidelity was calculated as described in Refs. \cite{ghosh:prb17, suppl}. For the phonon noise, we considered the same pulse sequence. However, we modified the Ramsey sequence in Fig.~\ref{fig:L_t_dependence_BR} (b) by setting $t=0$ during the $Z$ rotation, as proposed in \cite{ghosh:prb17}. This has the added benefit of suppressing the phonon-induced tunneling during Ramsey oscillations, so that decoherence occurs only during Larmor precession. To compute the average gate fidelity for phonon-induced noise during the free evolution portion of the Larmor sequence, $F_{ph}$, we average over initial states using the standard definition $F_{ph}=1/4\sum_\xi\mbox{Tr}\{\sqrt{\sqrt{\rho_\xi^{id}}\rho_\xi\sqrt{\rho_\xi^{id}}}\}$. Here, $\xi$ are indices denoting the initial states that form a regular tetrahedron on the Bloch sphere \cite{bowdrey:pla02}, and $\rho_\xi^{id}$ is a density matrix for the ideal system, i.e. without phonons. The results of our calculations are shown in Fig.~\ref{fig:L_t_dependence_BR} (b) for both types of noise. For phonons, the infidelity increases with $|t|$, for the same reason as in Fig.~\ref{fig:L_t_dependence_BR} (a). For charge noise, the fidelity shows the opposite trend, due to the suppression of leakage and broadening of the sweet spot.  As a result, an optimal working point emerges near $t=15\ \mu$eV, corresponding to a maximum fidelity $> 99.99\%$. 

\begin{figure}[tb] 
\begin{center}
\includegraphics[width=0.9\linewidth]{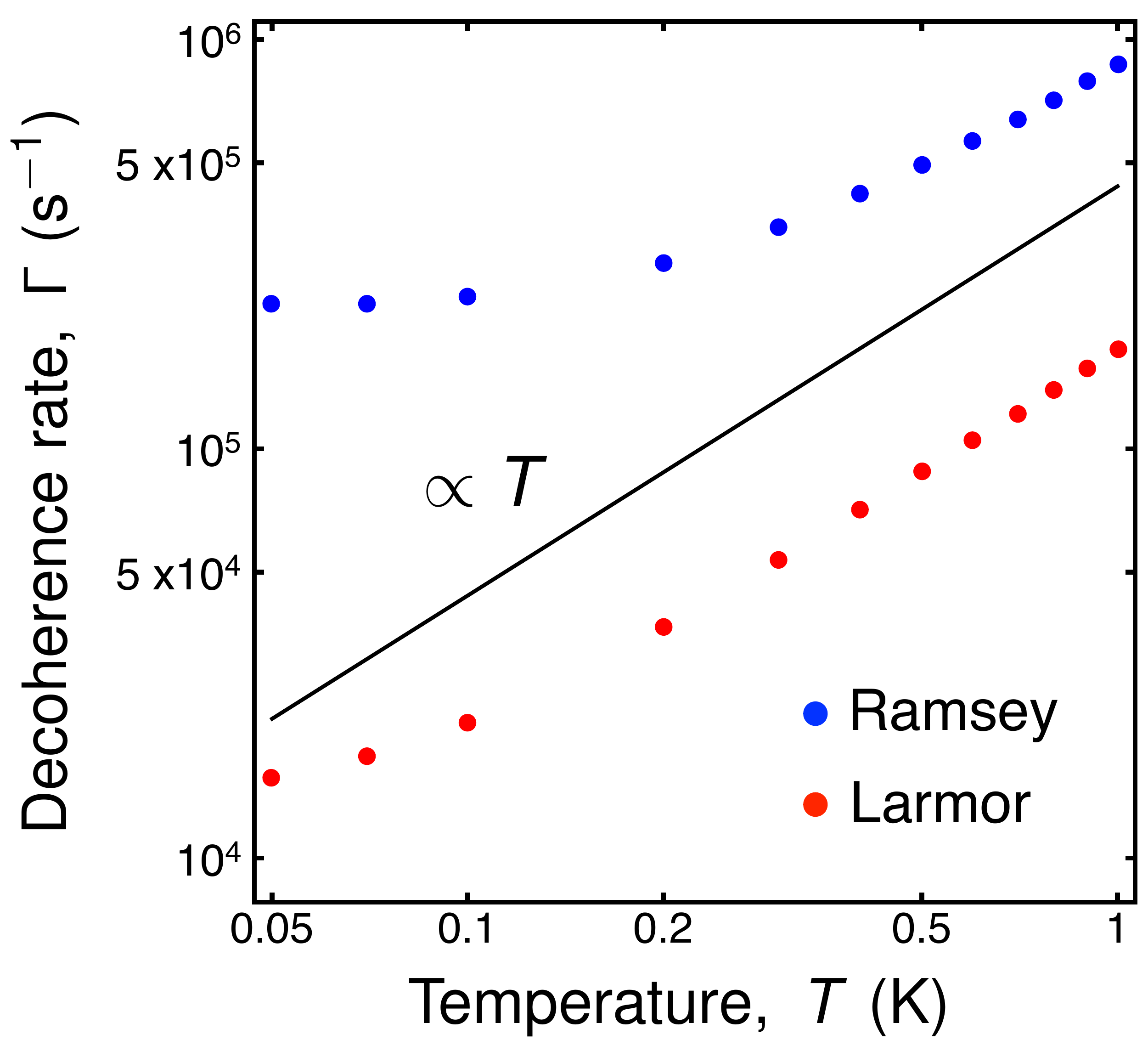}
\caption{Temperature dependence of the phonon-induced decoherence rates for Ramsey (blue dots) and Larmor (red dots) pulse sequences, at $t=-8.2\ \mu$eV, $l=18.5$ nm, and $a=90$ nm. Black line serves as a guide to indicate the dependence proportional to temperature. The transition to the linear regime occurs at lower temperatures for the Larmor pulse sequence, because the relevant energy scales are smaller than those of the Ramsey sequence, as indicated in Fig.~\ref{fig:pulses_setup} (d). Consequently, a high-temperature expansion of Bose-Einstein distribution is valid at lower temperatures for the Larmor sequence, than for the Ramsey sequence.}
\label{fig:temperature_dependence_BR}
\end{center}
\end{figure} 

For phonon noise it is possible to tune the decoherence rates by modifying geometrical device parameters. The dependence of $\Gamma_{\rm Lar}$ and $\Gamma_{\rm Ram}$ on interdot distance is shown in Fig.~\ref{fig:interdot_dist_dependence_BR}. Here, the tunnel coupling $t=-8.2\ \mu$eV is held constant; experimentally this can be accomplished by varying gate voltages. We take $l=18.5$ nm for all points here, to consider solely effects of changing $a$. Fig.~\ref{fig:interdot_dist_dependence_BR} shows that the rates grow with $a$ as a linear combination of power laws $a^2$ and $a^4$, which is justified by expanding the functions $J_{1L}$ and $J_{01}$ in small values of their exponential arguments. For $\Gamma_{\rm Ram}$ this expansion is not strictly valid, however empirically, we find that a fit with the same power laws seems to work. The physics underlying the growth of $\Gamma_{\rm Lar}$ and $\Gamma_{\rm Ram}$ with $a$ is that the decoherence is dominated by phonons with wavelengths larger than the triple dot size; as the QD separations are increased, the phonon causes larger shifts within the triple QD, and consequently more decoherence. 

The temperature dependence of the decoherence rates is shown in Fig.~\ref{fig:temperature_dependence_BR}. Both rates grow very slowly below $0.1$ K, particularly $\Gamma_{\rm Ram}$, and then increase linearly with $T$ at higher temperatures. The linear dependence on temperature appears when $\hbar\omega_{10},\hbar\omega_{1L}\ll k_BT$, causing the Bose-Einstein distribution to reduce to a classical distribution. The transition occurs sooner for $\Gamma_{\rm Lar}$ than for $\Gamma_{\rm Ram}$, because the Ramsey pulse sequence involves larger energy splittings. 

In conclusion, we have studied phonon-induced decoherence of a charge quadrupole qubit to determine its optimal working regime. We find that decoherence rates grow quickly with energy level splitting, due to the strong dependence of the phonon density of states and electron-phonon interaction matrix elements on it. This suggests that large tunnel couplings and quadrupolar detunings should be avoided during qubit operation. However, decoherence caused by charge noise is found to decrease rapidly with tunnel coupling, due to the suppression of leakage and the broadening of the sweet spot. Hence an optimal working point emerges for the tunnel coupling, which we estimate to be $t\simeq15\ \mu$eV for typical devices, corresponding to average gate fidelities greater than $99.99\%$, with X gate periods of $0.15$ ns and Z gate periods of $0.06$ ns. We also find that smaller dot separations tend to reduce the phonon-induced decoherence; a similar trend was previously observed for the coupling to charge noise \cite{friesen:ncom17}.

We thank Mark Eriksson, Joydip Ghosh, and Zhenyi Qi for helpful discussions. This work was supported in part by ARO (W911NF-15-1-0248, W911NF-17-1-0274) and the Vannevar Bush Faculty Fellowship program sponsored by the Basic Research Office of the Assistant Secretary of Defense for Research and Engineering and funded by the Office of Naval Research through Grant No. N00014-15-1-0029.  The views and conclusions contained here are those of the authors and should not be interpreted as representing the official policies, either expressed or implied, of the Army Research Office (ARO), or the U.S. Government.

\newpage
\renewcommand{\theequation}{S\arabic{equation}}
\setcounter{equation}{0}
\renewcommand{\thefigure}{S\arabic{figure}}
\renewcommand{\figurename}{Supplementary Fig.}

\setcounter{figure}{0}
\renewcommand{\thesection}{S\arabic{section}}
\setcounter{section}{0}
\begin{widetext}
\section*{Supplemental Material}
In this Supplemental Material we present additional details and calculations regarding phonon-induced decoherence of a charge quadrupole qubit. The Supplemental Material is organized as follows. In Sec. \ref{app:wavefunction} we present the wave function of an electron in the heterostructure growth direction. The derivation of the expression for the tunnel coupling is shown in Sec. \ref{sec:TunnelCoupling}. We give details regarding electron-phonon interaction Hamiltonian in Si in Sec. \ref{sec:TheModel}. In Sec. \ref{sec:BlochRedfieldTheory} we show the derivation of the equation for the reduced density matrix of the system interacting with bath. We use this equation to receive the results presented in the main text. In Sec. \ref{sec:FermiGoldenRule} we present our calculation of the transition rates via Fermi Golden rule. As these rates describe particular relaxation processes, they explain results presented in the main text in more detail. Sec. \ref{sec:Fidelity} explains how we calculated charge noise-induced infidelity of the qubit.

\section{The wave function of an electron in the heterostructure growth direction}
\label{app:wavefunction}
The full wave function of an electron in a triple quantum dot can be represented as a product of a component in the direction of heterostructure growth ($z$) and a lateral component ($x-y$), because the Hamiltonian $H_0$ can be divided into a sum of $z$ and $x-y$ terms. The lateral components of the wave functions $\psi(x,y)$ were discussed in the main text. Here with summarize our treatment of the remaining component $\varphi(z)$. A convenient approximate expression for the wave function of an electron in the direction of quantum well confinement is \cite{gamble:prb12}
\begin{equation}
\varphi(z)=\frac{\sqrt{2}i\sin{k_0z}}{\pi^{1/4}\sqrt{d(1-e^{-k_0^2d^2})}}\exp{\left[-\frac{z^2}{2d^2}\right]},
\end{equation} 
where $k_0$ is a wave vector corresponding to the minima of the conduction band in Si, $k_0\approx 0.82\times 2\pi/a_0$, with the length of the Si cubic cell $a_0=0.54$ nm. The parameter $d$ defines the width of the quantum well. For our calculations we simplified $\varphi$ as follows:
\begin{equation}
\varphi(z)\approx\frac{1}{\pi^{1/4}\sqrt{d}}\exp{\left[-\frac{z^2}{2d^2}\right]}.
\end{equation} 
Such simplification does not affect the results we get via Fermi Golden rule calculation noticeably, therefore we believe that it also does not affect the results of the Bloch-Redfield formalism. 
\section{Tunnel coupling}
\label{sec:TunnelCoupling}
From Eq. (1) in the main text, the definition of tunnel coupling is $t=\langle\Phi_R|H_0|\Phi_C\rangle$. Therefore we evaluate $t$ considering tunneling of a particle in a double-well potential $\frac{\hbar^2}{2m_{\rm eff}l^4a^2}(x^2-\frac{a^2}{4})^2$. This potential includes only two quantum wells, but this is sufficient for our purposes because the third dot is far away, and its effect is exponentially suppressed. To evaluate $t$ we use the wave functions constructed similar to $\Phi_R$ and $\Phi_C$, namely \cite{burkard:prb99, mattis:book81}
\begin{eqnarray}
\tilde{\Phi}_{L,R}&=&\frac{\psi^{a/2}_{L,R}-\tilde{g}\psi^{a/2}_{R,L}}{\sqrt{1-2\tilde{g}\tilde{S}+\tilde{g}^2}},\\
\tilde{S}&=&\langle\psi_L^{a/2}|\psi_R^{a/2}\rangle,\\
\tilde{g}&=&\frac{1-\sqrt{1-\tilde{S}^2}}{\tilde{S}},
\end{eqnarray}
where $\psi^{a/2}_{L,R}$ are the same as for the $x$-components of $\psi_{R,L}$ in the main text, but shifted along $x$ by $a/2$, not $a$. The expression for $t$ then reads
\begin{equation}
t=-\frac{3\hbar^2(a^2+4l^2)}{64m_{\rm eff}l^4\sinh{\frac{a^2}{4l^2}}}.
\end{equation}
The rapid growth of $t$ with $l$ arises because $t\propto \sinh^{-1}{\frac{a^2}{4l^2}}$. 

\section{The electron-phonon interaction Hamiltonian}
\label{sec:TheModel}
In this Section we present a detailed expression for the electron-phonon Hamiltonian in bulk Si; it is given by \cite{yu:book10}
\begin{equation}
\label{eq:ElPhHamiltonian}
H_{el-ph}=\Xi_d \mbox{Tr}{\bm \varepsilon}+\Xi_u \varepsilon_{zz},
\end{equation}
where ${\bm \varepsilon}$ is a strain tensor defined as \cite{kornich:arxiv16, kornich:prb14}
\begin{eqnarray} 
\varepsilon_{ij}&=&\frac{1}{2}\left(\frac{\partial u_i}{\partial r_j}+\frac{\partial u_j}{\partial r_i}\right),\\
\bm{u}&=&\sum_{q,s}\sqrt{\frac{\hbar}{2\rho_{Si} V q v_s}}\bm{e}_{\bm{q}s}(b_{\bm{q},s}\mp_sb_{-\bm{q},s}^\dagger)e^{i\bm{q}\cdot\bm{r}},
\end{eqnarray}
where $\Xi_d$ and $\Xi_u$ are deformation potential constants, ${\bm u}$ is a displacement operator, $s\in \{l,t_1,t_2\}$ (corresponding to one longitudinal and two transverse acoustic modes), $v_s$ is the speed of sound for each of these modes, $\rho_{Si}$ is the mass density of bulk Si, $V$ is the volume we consider, $b_{\bm{q},s}$ and $b_{\bm{q},s}^\dagger$ are annihilation and creation operators for phonons of mode $s$ and wave vector $\bm{q}$. The polarization vectors are chosen as follows: $\bm{e}_{\bm{q},l}=\bm{q}/q$, $\bm{e}_{-\bm{q},t_1}=-\bm{e}_{\bm{q},t_1}$ and $\bm{e}_{-\bm{q},t_2}=\bm{e}_{\bm{q},t_2}$ leading to the definition $\mp_l=-$, $\mp_{t_1}=-$, and $\mp_{t_2}=+$. The explicit expressions for polarizations are
\begin{eqnarray}
\bm{e}_{\bm{q},l}&=&\begin{pmatrix}
\cos{\phi_q}\sin{\theta_q}\\
\sin{\phi_q}\sin{\theta_q}\\
\cos{\theta_q}
\end{pmatrix},\\
\bm{e}_{\bm{q},t_1}&=&\begin{pmatrix}\sin{\phi_q}\\ -\cos{\phi_q}\\ 0\end{pmatrix},\\
\bm{e}_{\bm{q},t_2}&=&\begin{pmatrix}\cos{\phi_q}\cos{\theta_q}\\
\sin{\phi_q}\cos{\theta_q}\\
-\sin{\theta_q}\end{pmatrix},
\end{eqnarray}
where $0\leq\phi_q< 2\pi$ and $0\leq \theta_q< \pi$ are angles describing the phonon wavevector $\bm{q}$ in spherical coordinates $\bm{q}=\begin{pmatrix}q\cos{\phi_q}\sin{\theta_q}\\ q\sin{\phi_q}\sin{\theta_q}\\ q\cos{\theta_q} \end{pmatrix}$. Writing Eq. (\ref{eq:ElPhHamiltonian}) in a more explicit form, we get
\begin{equation} 
H_{el-ph}=i\Xi_d\sum_{\bm{q}}\sqrt{\frac{\hbar q}{2\rho_{Si} V v_l}}(b_{\bm{q},l}-b_{-\bm{q},l}^\dagger)e^{i\bm{q}\cdot \bm{r}}+i\Xi_u\sum_{q,s}\sqrt{\frac{\hbar}{2\rho_{Si} V q v_s}}q_ze^z_{\bm{q}s}(b_{\bm{q},s}\mp_sb_{-\bm{q},s}^\dagger)e^{i\bm{q}\cdot\bm{r}}.
\end{equation}

\begin{figure}[tb]
\begin{center}
\includegraphics[width=0.45\linewidth]{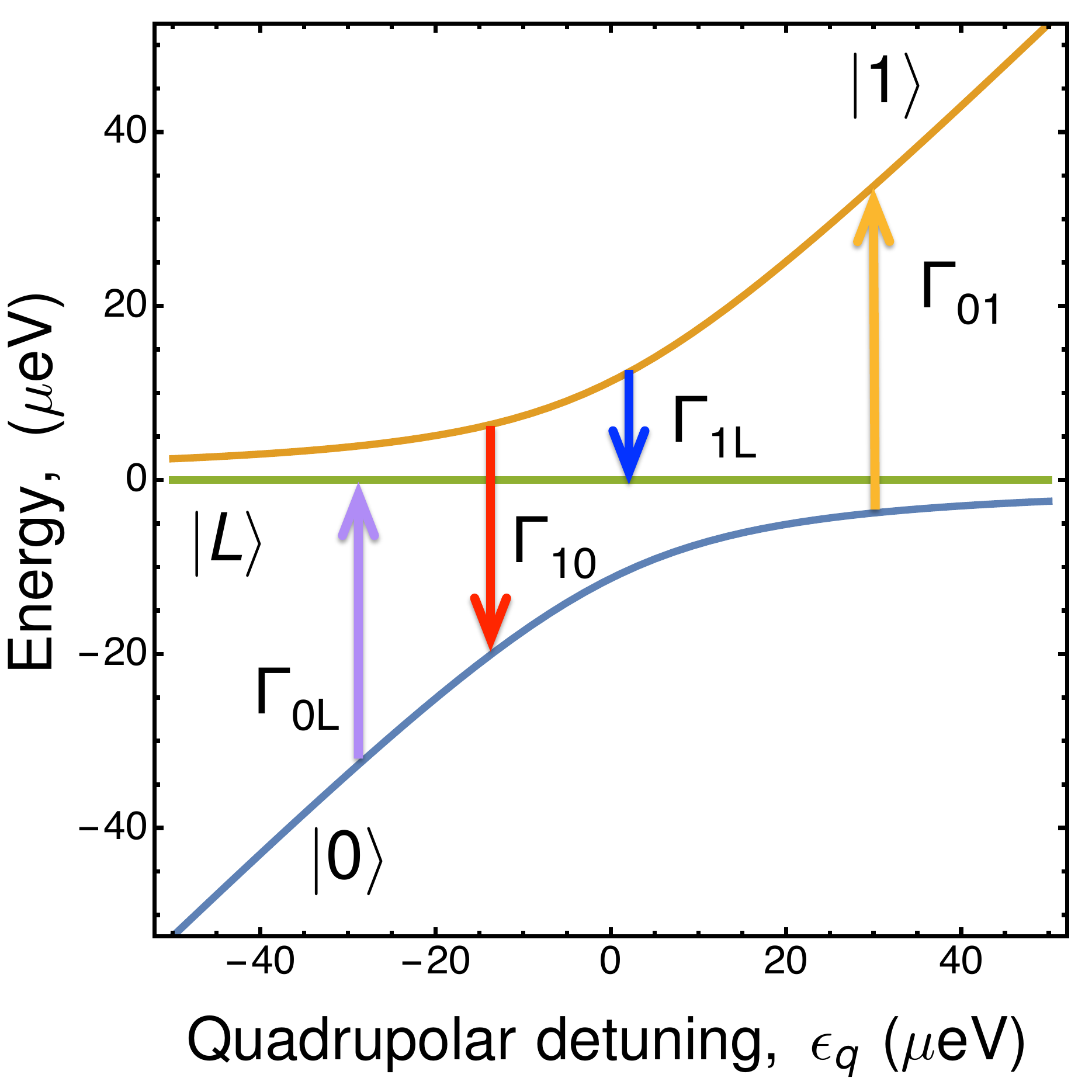}
\caption{Dependence of the energies of $|0\rangle$, $|1\rangle$, and $|L\rangle$ states on quadrupolar detuning $\epsilon_q$. Arrows show all possible one-phonon relaxation processes, where $\Gamma_{10}$ and $\Gamma_{1L}$ are found to be dominant. }
\label{fig:spectrum_cq_qubit}
\end{center}
\end{figure}

\begin{figure}[tb]
\begin{center}
\includegraphics[width=0.5\linewidth]{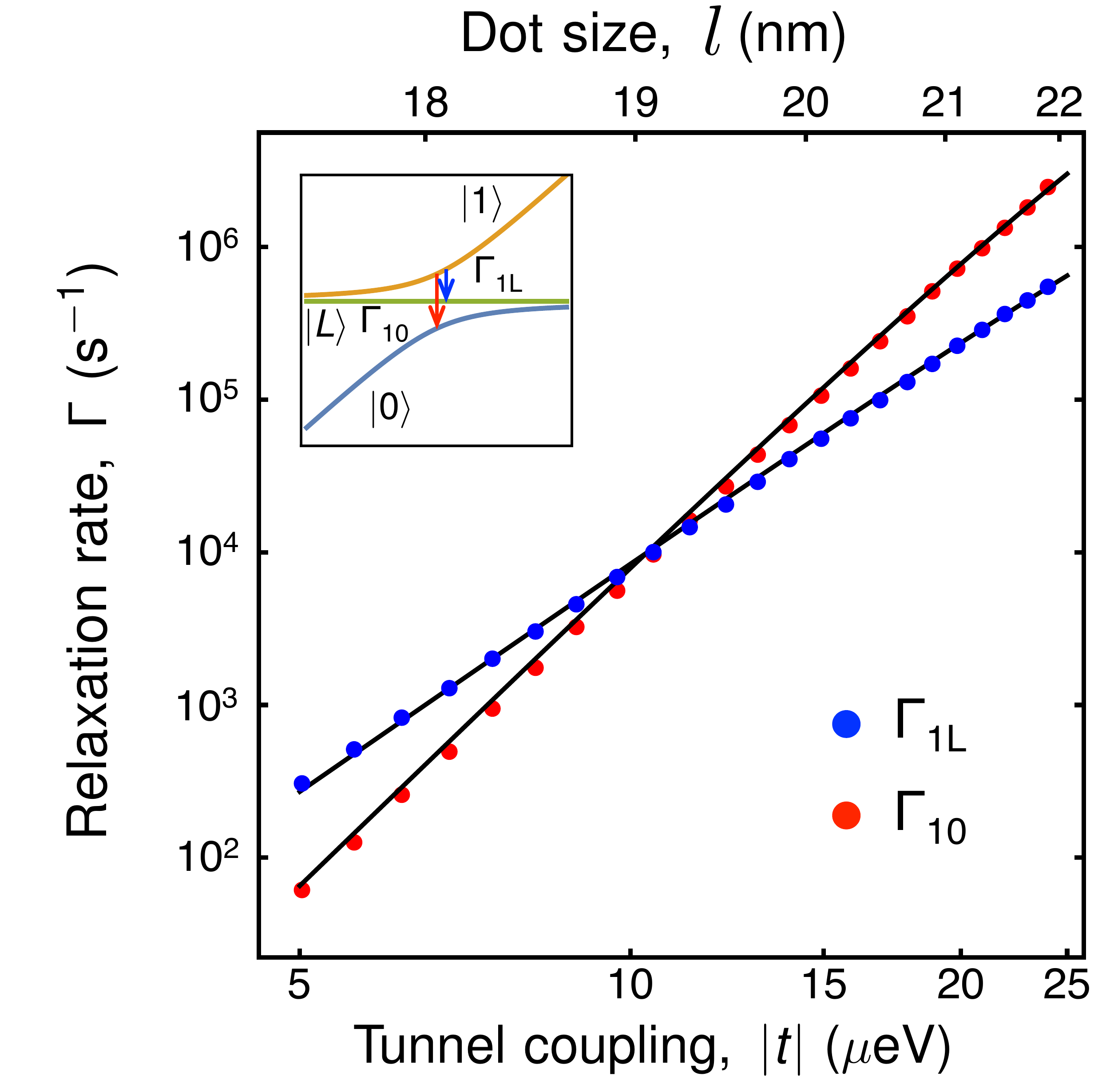}
\caption{Dependence of the relaxation rates $\Gamma_{10}$ and $\Gamma_{1L}$ at $\epsilon_q=0$ on the absolute value of tunnel coupling $|t|$. The values of the QD size, $l$, corresponding to the values of $|t|$ are given on the top axis. The inset shows the relaxation processes that give rise to $\Gamma_{10}$ and $\Gamma_{1L}$ on the energy level diagram. Both relaxation rates increase strongly with $|t|$. Here, interdot separation $a=90$ nm, and temperature $T=50$ mK.}
\label{fig:L_t_dependence_a_90}
\end{center}
\end{figure}

\begin{figure}[tb]
\begin{center}
\includegraphics[width=0.5\linewidth]{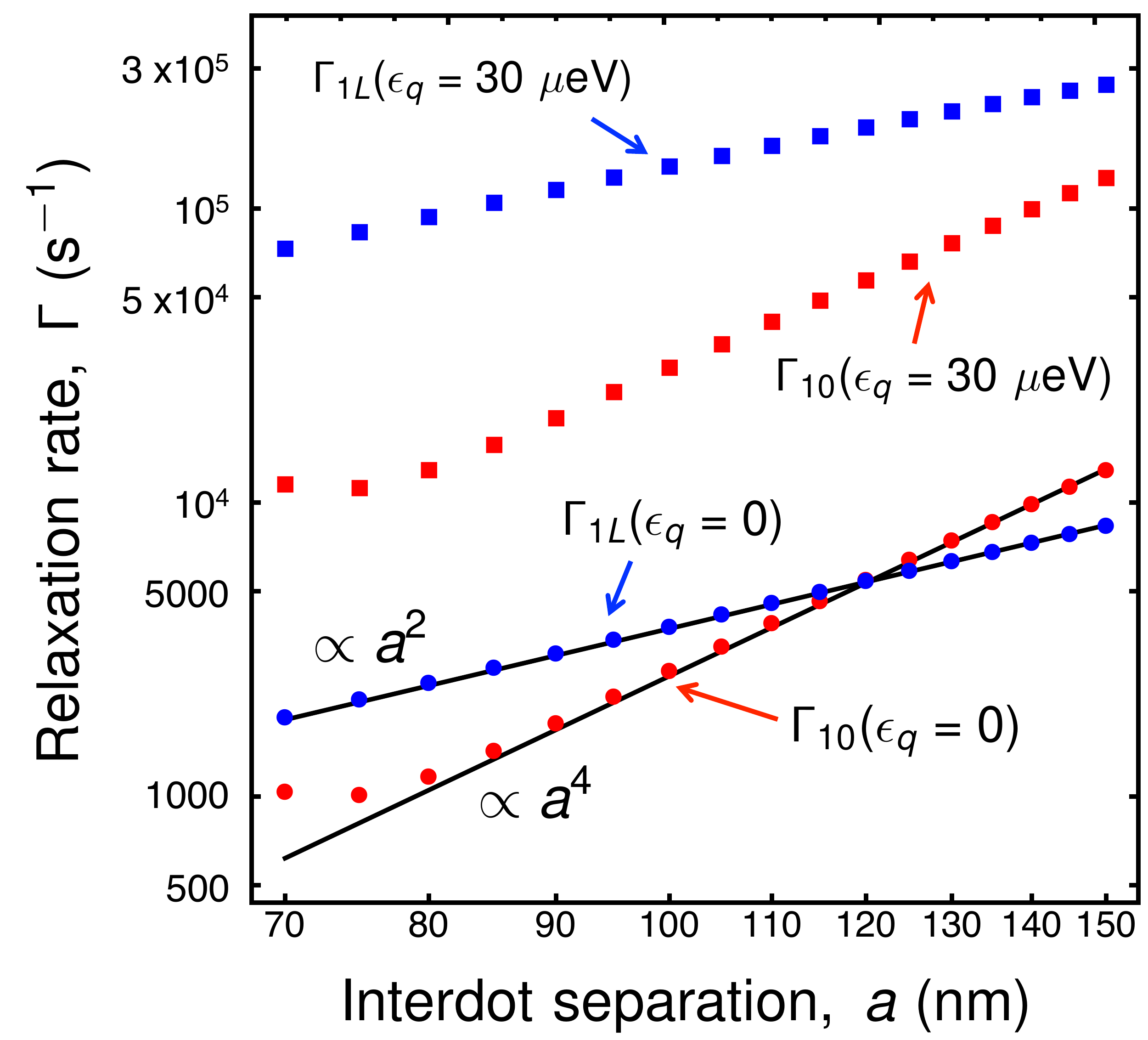}
\caption{Dependence of the relaxation rates $\Gamma_{10}$ and $\Gamma_{1L}$ on the interdot separation $a$ for quadrupolar detuning $\epsilon_q=0$ and $\epsilon_q=30\ \mu$eV. In general, the rates grow with $a$ because as the distance between the dots increases, the phase difference of the phonon wave in the different dots increases. Consequently, the phonons give rise to larger distortions of the triple QD potential. Here, the size of the dots is $l=18.5$ nm, tunnel coupling $t=-8.2\ \mu$eV, and temperature $T=50$ mK.}
\label{fig:interdot_dist_dependence_t_-8.2}
\end{center}
\end{figure}

\begin{figure}[tb]
\begin{center}
\includegraphics[width=0.5\linewidth]{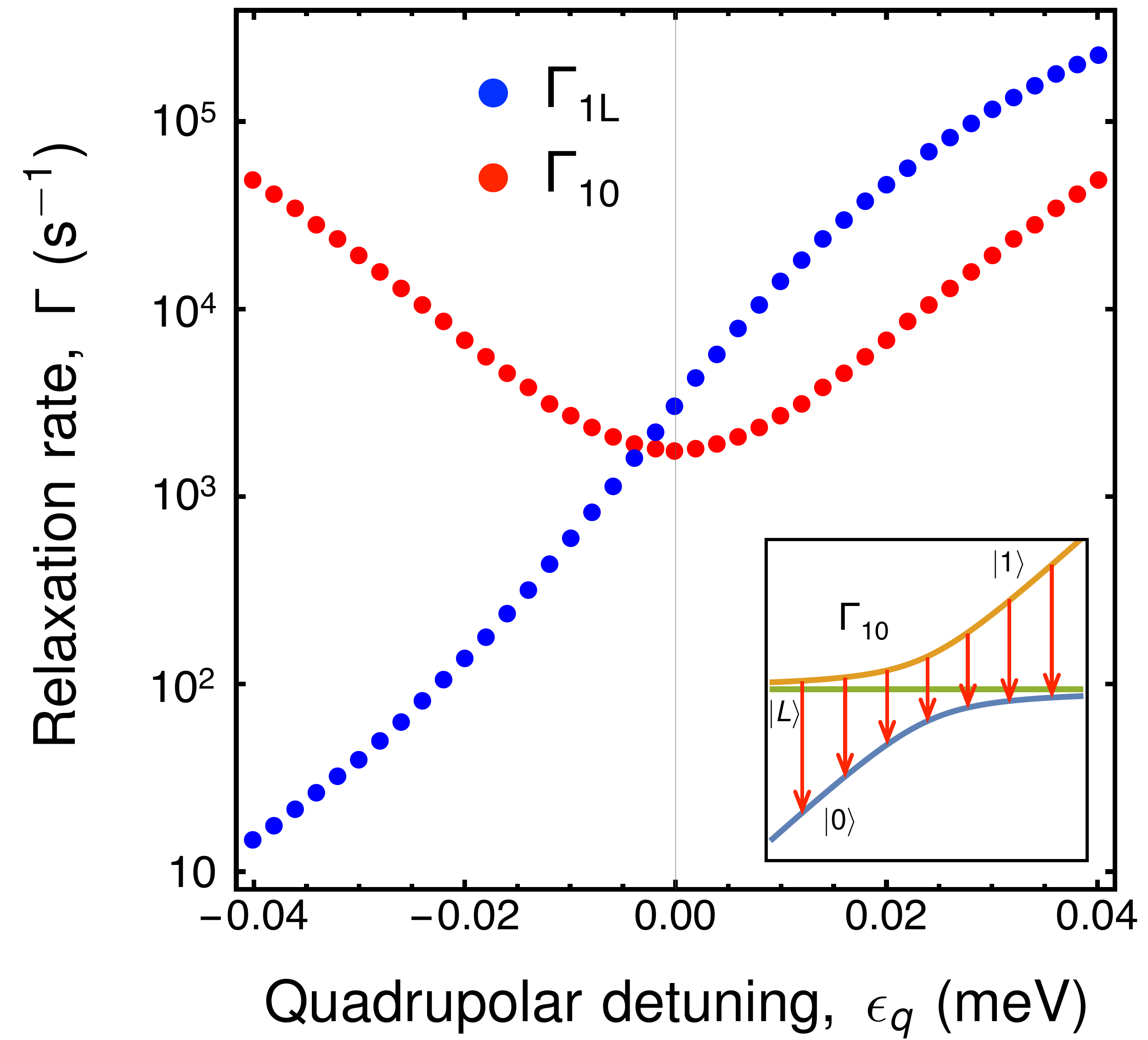}
\caption{Dependence of the relaxation rates $\Gamma_{10}$ and $\Gamma_{1L}$ on the quadrupolar detuning $\epsilon_q$. The dependence of $\Gamma_{10}$ exhibits symmetric behavior as the qubit energy splitting is also symmetric with respect to $\epsilon_q=0$. The splitting between $|1\rangle$ and $|L\rangle$ is very small when $\epsilon_q$ is large and negative; consequently $\Gamma_{1L}$ is very small in this regime. For qubit operation, both relaxation rates should be small, so operation near $\epsilon_q=0$ is expected to yield the the best results. Here, the set of parameters is the same as for Fig.~\ref{fig:interdot_dist_dependence_t_-8.2} and interdot separation $a=90$ nm.}
\label{fig:detuning_dependence_t_-8.2}
\end{center}
\end{figure}

\begin{figure}[tb]
\begin{center}
\includegraphics[width=0.5\linewidth]{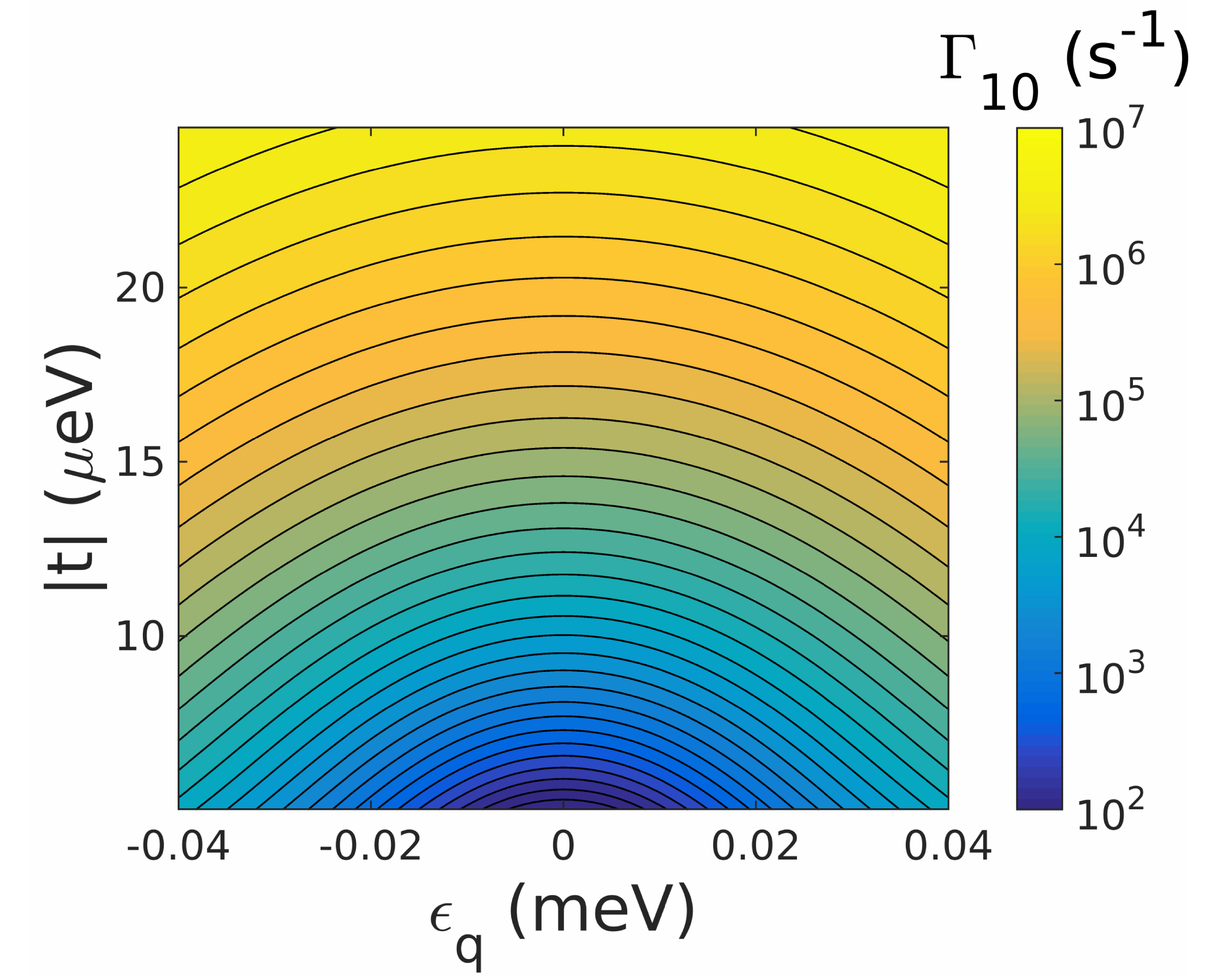}
\caption{Dependence of the relaxation rate $\Gamma_{10}$ on the quadrupolar detuning $\epsilon_q$ and tunnel coupling magnitude $|t|$, with symmetric behavior similar to Fig.~\ref{fig:detuning_dependence_t_-8.2}. Here, interdot separation $a=90$, temperature $T=50$ mK, and $l$ changes accordingly with $t$.}
\label{fig:contour_plot_qubit_states}
\end{center}
\end{figure}

\begin{figure}[tb]
\begin{center}
\includegraphics[width=0.5\linewidth]{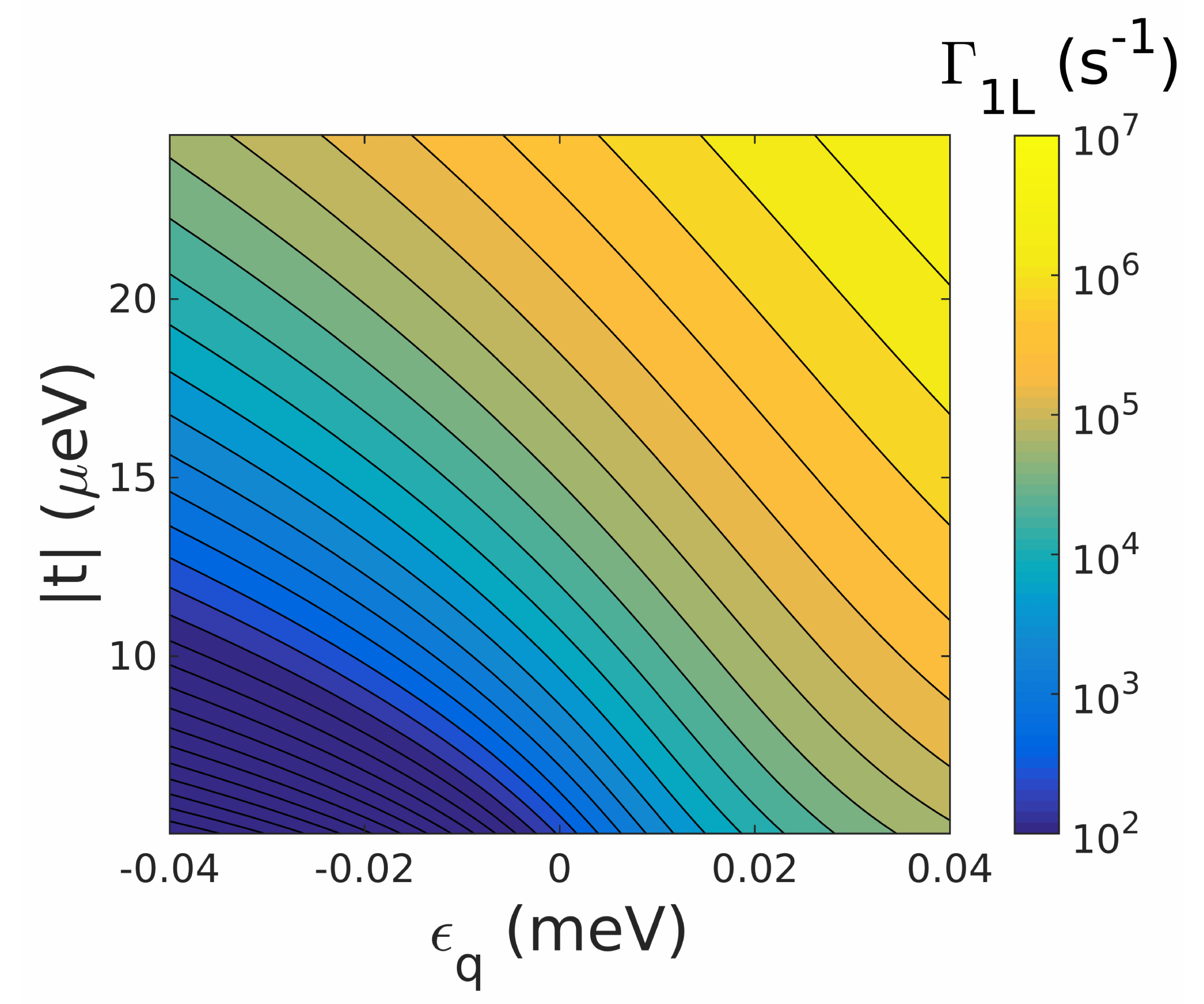}
\caption{Dependence of the relaxation rate $\Gamma_{1L}$ on the quadrupolar detuning $\epsilon_q$ and tunnel coupling magnitude $|t|$, with asymmetric behavior similar to Fig.~\ref{fig:detuning_dependence_t_-8.2}. The parameters used here are the same as for Fig.~\ref{fig:contour_plot_qubit_states}.}
\label{fig:contourplot_qubit_leakage}
\end{center}
\end{figure}

As later the $\tilde{H}_{el-ph}=U_d^{-1}H_{el-ph}U_d$ matrix elements will be important, particularly $\langle L|\tilde{H}_{el-ph}|1\rangle$ and $\langle1|\tilde{H}_{el-ph}|0\rangle$, we present them here
\begin{eqnarray}
\label{eq:LtildeH1}
&&\langle L|\tilde{H}_{el-ph}|1\rangle=\left[2(P_{RR}-P_{LL})-\frac{1}{t}(\epsilon_q+\sqrt{8t^2+\epsilon_q^2})(P_{LC}-P_{CR})\right]\sqrt{\frac{\sqrt{8t^2+\epsilon_q^2}-\epsilon_q}{2\sqrt{8t^2+\epsilon_q^2}}},\\
\label{eq:0tildeH1}
&&\langle0|\tilde{H}_{el-ph}|1\rangle=\frac{\sqrt{\sqrt{8t^2+\epsilon_q^2}-\epsilon_q}}{4t\sqrt{8t^2+\epsilon_q^2}\sqrt{\sqrt{8t^2+\epsilon_q^2}+\epsilon_q}}[t(\epsilon_q+\sqrt{8t^2+\epsilon_q^2})(P_{LL}+P_{RR}-2P_{CC})+\\ \nonumber &&+4t^2(P_{CR}^\dagger+P_{LC}-P_{LC}^\dagger-P_{CR})+(P_{CR}^\dagger+P_{LC})\epsilon_q(\epsilon_q+\sqrt{8t^2+\epsilon_q^2})].
\end{eqnarray} 

\section{Bloch-Redfield Formalism}
\label{sec:BlochRedfieldTheory}

Here we derive the equation for the reduced density matrix that describes our three-level system, which we use to obtain decoherence rates from the decay of the occupation of the system's states. We follow the standard procedure for deriving Bloch equations \cite{blum:book96, golovach:prl04, borhani:prb06}. 

The Hamiltonian in our problem is given by
\begin{equation}
H=H_q+H_{ph}+H_{q-ph},
\end{equation}
where $H_q$ is the Hamiltonian of a qubit (including the leakage state in our case), $H_{ph}$ describes the Hamiltonian of the phonons and $H_{q-ph}$ is an interaction between them. The von Neumann equation for a full density matrix in the Schr\"odinger representation is 
\begin{equation}
\dot{\tilde{\rho}}=i[\tilde{\rho}, H],
\end{equation}
where we set $\hbar = 1$ for brevity. In the interaction representation defined as $\tilde{\rho}_{int}=e^{i (H_q+H_{ph})t}\tilde{\rho} e^{-i(H_q+H_{ph})t}$, the von Neumann equation becomes
\begin{equation}
\label{eq:rhoint}
\dot{\tilde{\rho}}_{int}=i[\tilde{\rho}_{int},H_{q-ph}^{int}],
\end{equation}
and integration yields
\begin{equation}
\tilde{\rho}_{int}(t)=i\int_0^td\tau[\tilde{\rho}_{int}(\tau),H_{q-ph}^{int}(\tau)]+\tilde{\rho}_{int}(0).
\end{equation}
Substituting into Eq. (\ref{eq:rhoint}) yields
\begin{equation}
\dot{\tilde{\rho}}_{int}(t)=-\int_0^td\tau[[\tilde{\rho}_{int}(\tau),H_{q-ph}^{int}(\tau)],H_{q-ph}^{int}(t)]+i[\tilde{\rho}_{int}(0),H_{q-ph}^{int}(t)].
\end{equation}
As we are mainly interested in the three-level system of the qubit with the leakage state, we trace over phonon degrees of freedom to obtain the reduced density matrix:
\begin{equation}
\label{eq:rhointintegral}
\dot{\tilde{\rho}}_{int}^{red}(t)=-\int_0^td\tau Tr_{ph}\{[[\tilde{\rho}_{int}(\tau),H_{q-ph}^{int}(\tau)],H_{q-ph}^{int}(t)]\}+iTr_{ph}\{[\tilde{\rho}_{int}(0),H_{q-ph}^{int}(t)]\}.
\end{equation}
We then apply the Born approximation: $\tilde{\rho}_{int}(t)=\rho_{int}^{ph}(0)\otimes \rho_{int}^q(t)$ and the Markov approximation $\tilde{\rho}_{int}(\tau)\rightarrow \tilde{\rho}_{int}(t)$ inside the integral. Here we note that $Tr_{ph}\{[\rho_{int}^{ph}(0)\otimes\rho_{int}^q(0),H_{q-ph}^{int}(t)]\}=0$ because $Tr_{ph}\{\rho_{int}^{ph} b_k\}=0$, and similarly for $a^\dagger_k$.  Therefore only the first term remains in Eq. (\ref{eq:rhointintegral}), which after the Born-Markov approximations reads
\begin{equation}
\dot{\rho}_{int}^{q}(t)=-\int_0^td\tau Tr_{ph}\{[[\rho_{int}^{ph}\otimes\rho_{int}^q(t),H_{q-ph}^{int}(\tau)],H_{q-ph}^{int}(t)]\}.
\end{equation}
We expand the commutator to obtain
\begin{eqnarray}
\dot{\rho}_{int}^{q}(t)=-\int_0^td\tau Tr_{ph}\{\rho_{int}^{ph}\rho_{int}^q(t)H_{q-ph}^{int}(\tau)H_{q-ph}^{int}(t)-H_{q-ph}^{int}(\tau)\rho_{int}^{ph}\rho_{int}^q(t)H_{q-ph}^{int}(t)\\ \nonumber-H_{q-ph}^{int}(t)\rho_{int}^{ph}\rho_{int}^q(t)H_{q-ph}^{int}(\tau)+H_{q-ph}^{int}(t)H_{q-ph}^{int}(\tau)\rho_{int}^{ph}\rho_{int}^q(t)\}.
\end{eqnarray}
We then project this equation onto the qubit-leakage basis (dropping the notation `int' for brevity), yielding:
\begin{eqnarray}
\label{eq:rhointexpanded}
\dot{\rho}_{nm}^{q}(t)=-\int_0^td\tau Tr_{ph}\{\rho^{ph}\rho_{nk}^q(t)H_{q-ph}^{kj}(\tau)H_{q-ph}^{jm}(t)-H_{q-ph}^{nk}(\tau)\rho^{ph}\rho_{kj}^q(t)H_{q-ph}^{jm}(t)-\\ \nonumber-H_{q-ph}^{nk}(t)\rho^{ph}\rho_{kj}^q(t)H_{q-ph}^{jm}(\tau)+H_{q-ph}^{nk}(t)H_{q-ph}^{kj}(\tau)\rho^{ph}\rho_{jm}^q(t)\},
\end{eqnarray}
where the convention of summing over repeated indices is used. We also switch to a different interaction picture in the following way:
\begin{equation}
H_{q-ph}^{nk}=\langle n|H_{q-ph}|k\rangle=\langle n|e^{i(H_q+H_{ph})t}H_{q-ph}e^{-i(H_q+H_{ph})t}|k\rangle=e^{i\omega_{nk}t}e^{iH_{ph}t}\langle n|H_{q-ph}|k\rangle e^{-iH_{ph}t}=e^{i\omega_{nk}t}\bar{H}_{q-ph}^{nk},
\end{equation}
where the bar denotes the interaction representation with respect to the phonon Hamiltonian only, and $\omega_{nk}=E_n-E_k$, where $E_n$ and $E_k$ are the energies of states $n$ and $k$ respectively. We apply such representation to all the Hamiltonian terms in Eq. (\ref{eq:rhointexpanded}), obtaining
\begin{eqnarray}
\dot{\rho}_{nm}^{q}(t)=-\int_0^td\tau Tr_{ph}\{\rho^{ph}\rho_{nk}^q(t)\bar{H}_{q-ph}^{kj}(\tau)\bar{H}_{q-ph}^{jm}(t)e^{i\omega_{kj}\tau+i\omega_{jm}t}-\bar{H}_{q-ph}^{nk}(\tau)\rho^{ph}\rho_{kj}^q(t)\bar{H}_{q-ph}^{jm}(t)e^{i\omega_{nk}\tau+i\omega_{jm}t}\\ \nonumber-\bar{H}_{q-ph}^{nk}(t)\rho^{ph}\rho_{kj}^q(t)\bar{H}_{q-ph}^{jm}(\tau)e^{i\omega_{nk}t+i\omega_{jm}\tau}+\bar{H}_{q-ph}^{nk}(t)\bar{H}_{q-ph}^{kj}(\tau)\rho^{ph}\rho_{jm}^q(t)e^{i\omega_{nk}t+i\omega_{kj}\tau}\}.
\end{eqnarray}
We make the substitution $\tau=t-\tau'$, obtaining
\begin{eqnarray}
&&\dot{\rho}_{nm}^{q}(t)=\\ \nonumber&&-\int_0^td\tau' Tr_{ph}\{\bar{H}_{q-ph}^{kj}(-\tau')\bar{H}_{q-ph}^{jm}(0)\rho^{ph}\rho_{nk}^q(t)e^{i(\omega_{kj}+\omega_{jm})t-i\omega_{kj}\tau'}-\bar{H}_{q-ph}^{jm}(\tau')\bar{H}_{q-ph}^{nk}(0)\rho^{ph}\rho_{kj}^q(t)e^{i(\omega_{nk}+\omega_{jm})t-i\omega_{nk}\tau'} \ \ \ \ \ \ \\ \nonumber&&-\bar{H}_{q-ph}^{jm}(-\tau')\bar{H}_{q-ph}^{nk}(0)\rho^{ph}\rho_{kj}^q(t)e^{i(\omega_{nk}+\omega_{jm})t-i\omega_{jm}\tau'}+\bar{H}_{q-ph}^{nk}(\tau')\bar{H}_{q-ph}^{kj}(0)\rho^{ph}\rho_{jm}^q(t)e^{i(\omega_{nk}+\omega_{kj})t-i\omega_{kj}\tau'}\}.\end{eqnarray}  
Defining
\begin{eqnarray}
\label{eq:GammaPlus}
\Gamma_{jmnk}^+&=&\int_0^td\tau' Tr_{ph}\{\bar{H}_{q-ph}^{jm}(\tau')\bar{H}_{q-ph}^{nk}(0)\rho^{ph}\}e^{-i\omega_{nk}\tau'},\\
\label{eq:GammaMinus}
\Gamma_{jmnk}^-&=&\int_0^td\tau' Tr_{ph}\{\bar{H}_{q-ph}^{jm}(-\tau')\bar{H}_{q-ph}^{nk}(0)\rho^{ph}\}e^{-i\omega_{jm}\tau'}.
\end{eqnarray}
Our equation takes the form
\begin{eqnarray}
&&\dot{\rho}_{nm}^{q}(t)= \nonumber\\
&&=-\Gamma_{kjjm}^-\rho_{nk}^q(t)e^{i(\omega_{kj}+\omega_{jm})t}+\Gamma_{jmnk}^+\rho_{kj}^q(t)e^{i(\omega_{nk}+\omega_{jm})t}+\Gamma_{jmnk}^-\rho_{kj}^q(t)e^{i(\omega_{nk}+\omega_{jm})t}-\Gamma_{nkkj}^+\rho_{jm}^q(t)e^{i(\omega_{nk}+\omega_{kj})t}.\ \ \ \ 
\end{eqnarray}
Note, that we omitted the superscript `int' here to simplify the notation; however the reduced density matrix is in interaction representation as shown above Eq. (\ref{eq:rhoint}). To transform back to the Schr\"odinger representation we use
\begin{eqnarray}
\rho_{nm}^{int,q}(t)&=&e^{i\omega_{nm}t}\rho_{nm}^q(t),\\
\partial_t \rho_{nm}^{int,q}(t)&=&e^{i\omega_{nm}t}(i\omega_{nm}+\partial_t)\rho_{nm}^q(t).
\end{eqnarray}
Finally, in the Schr\"odinger representation, we obtain
\begin{eqnarray}
\dot{\rho}_{nm}^q(t)=-i\omega_{nm}\rho_{nm}^q(t)-\Gamma_{kjjm}^-\rho_{nk}^q(t)+(\Gamma_{jmnk}^++\Gamma_{jmnk}^-)\rho_{kj}^q(t)-\Gamma_{nkkj}^+\rho_{jm}^q(t).
\end{eqnarray}
Here, the initial conditions are defined by the desired pulse sequence, as discussed in the main text and Fig. 1 (d). By computing the density matrix as a function of time for a particular initial condition, we determine how the system decays, to extract decoherence by fitting the decay to the exponential form.

\section{Fermi Golden Rule calculation of transition rates}
\label{sec:FermiGoldenRule}
In this section we use the Fermi Golden rule to calculate the transition rates of the relaxation processes between different states. These rates have similar form to Eqs. (\ref{eq:GammaPlus}) and (\ref{eq:GammaMinus}) which helps to better understand the behavior of the results obtained from Bloch-Redfield formalism.  We assume that the initial state is one of the qubit states. 

The Fermi Golden rule expression for the transition rate from initial state $|i\rangle$ to final state $|f\rangle$ is 
\begin{equation}
\Gamma_{if}=\frac{2\pi}{\hbar}\sum_{{\bm q},s}|\langle f|\tilde{H}_{el-ph}^{{\bm q}, s}|i\rangle|^2\delta(E_i-E_f\pm\hbar\omega_{{\bm q},s}),
\end{equation}
where $-$ is for absorption and $+$ is for emission of a phonon, $\tilde{H}_{el-ph}=\sum_{{\bm q},s}\tilde{H}_{el-ph}^{{\bm q},s}$, and $|i\rangle$ and $|f\rangle$ are the initial and final states of our system respectively, including the electron and phonon bath. Here, the initial and final energies of the electron are $E_i$ and $E_f$, and the frequency $\omega_{{\bm q},s}$ corresponds to a phonon that is emitted or absorbed.
 
The full expressions for the rates are lengthy, therefore we do not include them all here. However we provide one of the shortest of the expressions as an example:
\begin{eqnarray}
\label{eq:Gamma1L}
&&\Gamma_{1L}=\sum_{s}\int \frac{d\theta_q d\phi_q \sin{\theta_q}}{(2\pi)^2\hbar^2v_s}\\ \nonumber&&\times\left|\frac{\left[2t[p_{RR,s}(q,\phi_q,\theta_q)-p_{LL,s}(q,\phi_q,\theta_q)]+[p_{CR,s}(q,\phi_q,\theta_q)\pm_s p_{LC,s}^*(q,\phi_q+\pi,\pi-\theta_q)]\left[\sqrt{8t^2+\epsilon_q^2}+\epsilon_q\right]\right]}{4\sqrt{2}t(8t^2+\epsilon_q^2)^{1/4}}\right|^2_{q=q_{1L}} \\ \nonumber 
&&\times\sqrt{\sqrt{8t^2+\epsilon_q^2}-\epsilon_q}\left[\frac{1}{2\hbar v_s}\left(\sqrt{8t^2+\epsilon_q^2}+\epsilon_q\right)\right]^2 \left[n_{BE}\left(\frac{1}{2}\left[\sqrt{8t^2+\epsilon_q^2}+\epsilon_q\right]\right)+1\right],
\end{eqnarray}
where $q_{1L}=\frac{\sqrt{8t^2+\epsilon_q^2}+\epsilon_q}{2\hbar v_s}$, $n_{BE}$ is the Bose-Einstein distribution function, and $p$ is defined in terms of the electron-phonon matrix elements as follows: 
\begin{eqnarray}
P_{ij}=\sum_{\bm{q}}p_{ij,l}(q,\phi_q,\theta_q)(b_{\bm{q},l}-b_{-\bm{q},l}^\dagger)+p_{ij,t_1}(q,\phi_q,\theta_q)(b_{\bm{q},t_1}-b_{-\bm{q},t_1}^\dagger)+p_{ij,t_2}(q,\phi_q,\theta_q)(b_{\bm{q},t_2}+b_{-\bm{q},t_2}^\dagger).
\end{eqnarray}

The relaxation rates are calculated for the following set of Si material parameters: deformation potential constants $\Xi_d=5$ eV, $\Xi_u=8.77$ eV, longitudinal speed of sound $v_l=9\times 10^3$ m/s, transverse speed of sound $v_{t_1}=v_{t_2}=5.4\times 10^3$ m/s, and mass density $\rho_{Si}=2.33\ \rm g/cm^3$. The quantum well in $z$ direction is characterized by $d=2$ nm. 

Since we consider the low temperature $T=0.05$ K, the absorption rates are much slower than the emission. Mathematically this follows from the fact that emission rates have a $n_{BE}+1$ term, as in the last bracket of Eq. (\ref{eq:Gamma1L}), while absorption rates only have $n_{BE}$. Therefore, since $n_{BE}$ becomes very small for low temperatures, emission rates dominate over absorption. Therefore, the dominant processes are the emission of a phonon between the qubit states, $\Gamma_{10}$, and between $|1\rangle$ and $|L\rangle$, $\Gamma_{1L}$, depicted in Fig.~\ref{fig:spectrum_cq_qubit}.

When $\epsilon_q=0$ the main contribution to $\Gamma_{1L}$ in the parameters ranges we consider, comes from the difference $P_{RR}-P_{LL}$, appearing in Eq. (\ref{eq:LtildeH1}). Similarly, for $\Gamma_{10}$, the main contribution comes from the difference $P_{LL}+P_{RR}-2P_{CC}$ (see Eq. (\ref{eq:0tildeH1})) almost for all parameters considered. Consequently, the behaviors of the rates $\Gamma_{1L}$ and $\Gamma_{10}$ for $\epsilon_q=0$ can be fit as follows 
\begin{eqnarray}
\label{eq:Gamma1Lt}
\Gamma_{1L}&\propto& J_{1L}=|t|^3\left|\exp{\left(-\frac{i\sqrt{2}t a}{\hbar v_{t_1}}\right)}-\exp{\left(\frac{i \sqrt{2}t a}{\hbar v_{t_1}}\right)}\right|^2,\\
\label{eq:Gamma10t}
\Gamma_{10}&\propto& J_{10}=|t|^3\left|\exp{\left(-\frac{i2\sqrt{2}t a}{\hbar v_{t_1}}\right)}+\exp{\left(\frac{i2\sqrt{2}t a}{\hbar v_{t_1}}\right)}-2\right|^2.
 \end{eqnarray}
Here, $|t|^2$ comes from the phonon density of states, another $|t|$ comes from the form of the electron-phonon interaction and exponents contain the phase shifts between matrix elements $P_{LL}$, $P_{CC}$, and $P_{RR}$ due to the interdot separation $a$.  The transverse phonons dominate for these parameters, therefore the approximation with only $v_{t_1}$ works well. Following the fact that the relaxation processes $|1\rangle\leftrightarrow|L\rangle$ and $|1\rangle\leftrightarrow|0\rangle$ are dominating in the decoherence rates for Larmor and Ramsey pulse sequences, we used $J_{1L}$ and $J_{10}$ in the main text to approximate the results for the dependences of the decoherence rates on $|t|$ and $a$.

 As the dots are gate-defined, the size of each dot can be varied. Here, the transverse size of the dots is characterized by $l$. Since the experimentally relevant quantity is the tunnel coupling, here we present the dependence of $\Gamma_{10}$ and $\Gamma_{1L}$ on $|t|$, whereas $t$ is calculated using geometrical parameters $a$ and $l$ of the triple quantum dot, see Fig.~\ref{fig:L_t_dependence_a_90}. When $\sqrt{2}t a/(\hbar s_2)<1$, we can expand the exponents in Eq. (\ref{eq:Gamma1Lt}), obtaining the power law $|t|^5$, whereas for Eq. (\ref{eq:Gamma10t}), when $2\sqrt{2}t a/(\hbar s_2)<1$, we get the power-law $|t|^7$.

Another geometrical parameter we consider is the interdot separation $a$. We plot the dependence of $\Gamma_{1L}$ and $\Gamma_{10}$ on $a$ in Fig.~\ref{fig:interdot_dist_dependence_t_-8.2} for $\epsilon_q=0$ and $\epsilon_q=30\ \mu$eV. We see that $\Gamma_{10}$ and $\Gamma_{1L}$ are both more than $\sim 10$ times smaller for $\epsilon_q=0$ than for $\epsilon_q=30\ \mu$eV. In the former case we find that the rates depend on interdot distance $a$ as $\Gamma_{1L}\propto a^2$ and $\Gamma_{10}\propto a^4$, with very good precision for all $a$ in $\Gamma_{1L}$ and above $a\approx 80$ nm for $\Gamma_{10}$. Such power laws can be understood from the exponents in Eqs. (\ref{eq:Gamma1Lt}) and (\ref{eq:Gamma10t}), as they can be expanded and consequently give the power laws $a^2$ and $a^4$. The behavior of $\Gamma_{10}$ for $a<80$ nm can be explained as follows. The overlap between wave functions becomes strong enough to cause the off-diagonal matrix elements $P_{LC}$, $P_{CR}$, $P_{LC}^\dagger$, $P_{CR}^\dagger$ contribute significantly to the relaxation rate, therefore modifying the power-law. When $\epsilon_q=30\ \mu$eV, the arguments of the exponents in Eqs. (\ref{eq:Gamma1Lt}) and (\ref{eq:Gamma10t}) are not small for part of the parameter range, so it is no longer valid to expand in them. 

The increase of the rates with $a$ can be explained physically as follows. Since the phonons that correspond to the considered transitions are long-wavelength, if the triple quantum dot has small dimensions, it experiences the phonon as an approximately uniform shift in energy. If the dots are further apart, the same phonon produces larger shifts between the dots, appearing in the $H_{el-ph}$ matrix elements, which consequently produces more relaxation.

The dependence of $\Gamma_{1L}$ and $\Gamma_{10}$ on quadrupolar detuning is determined by the energy level spacing between the states we consider. As Fig.~\ref{fig:detuning_dependence_t_-8.2} shows, the rate $\Gamma_{10}$ is at a minimum when $\epsilon_q=0$ and exhibits symmetric behavior with respect to $\epsilon_q=0$ axis, while $\Gamma_{1L}$ has an asymmetric dependence on $\epsilon_q$. Such behavior corresponds to the change of the energy splittings between the states, as shown in the inset of Fig.~\ref{fig:detuning_dependence_t_-8.2}. We also present in Figs.~\ref{fig:contour_plot_qubit_states} and \ref{fig:contourplot_qubit_leakage} the contour plots for the dependences of $\Gamma_{10}$ and $\Gamma_{1L}$ on $|t|$ and $\epsilon_q$.

\section{Charge noise-induced infidelity}
\label{sec:Fidelity}
In this section we present the model we used to calculate quasistatic charge noise for charge quadrupole qubit. For that we followed the procedure described in Ref.~\onlinecite{ghosh:prb17}. We first define 
\begin{eqnarray}
H_z=\frac{\epsilon_q}{2}\begin{pmatrix}1 & 0 & 0 \\
0 & -1 & 0\\
0 & 0 & -1\end{pmatrix}, \ 
H_x=\sqrt{2}t\begin{pmatrix}0 & 1 & 0 \\
1 & 0 & 0 \\
0 & 0 & 0\end{pmatrix}, \
H_{\rm leak}=\delta \epsilon_d\begin{pmatrix}0 & 0 & 0 \\
0 & 0 & 1\\
0 & 1 & 0 \end{pmatrix}.
\end{eqnarray}
 Here $\delta\epsilon_d$ represents the charge noise on the gates; we take it $\delta\epsilon_d=1\ \mu$eV \cite{thorgrimsson:qi17}. The operators of $Z$ and $X$ rotations with charge noise are defined as follows
 \begin{eqnarray}
 U_z(\epsilon_q,\delta\epsilon_d,\phi)&=&\exp{\left[-i[H_z(\epsilon_q)+H_{\rm leak}(\delta\epsilon_d)]\phi/\epsilon_q\right]},\\
 U_x(t,\delta\epsilon_d,\theta)&=&\exp{\left[-i[H_x(t)+H_{\rm leak}(\delta\epsilon_d)]\theta/(2\sqrt{2}t)\right]},
 \end{eqnarray}
 for the rotation angles $\phi$ and $\theta$, which define corresponding gate times as $\tau_z=\phi \hbar/\epsilon_q$ and $\tau_x=\theta\hbar/(2\sqrt{2}t)$. To construct the rotation $X(\pi)$ we build the following pulse sequence
\begin{equation}
R=U_z\left(\epsilon_q,\delta\epsilon_d,\frac{\phi}{2}\right)U_x(t,\delta\epsilon_d,\theta)U_z\left(-\epsilon_q,\delta\epsilon_d,-\frac{\phi}{2}\right),
\end{equation} 
and take $\phi=2\pi$, $\theta=3\pi$, and $\epsilon_q=-t\phi\cot{\left(\frac{\theta}{4}\right)}/\sqrt{2}$, which gives $Z(\pi)X(3\pi)Z(-\pi)$.
The definition of the average gate fidelity we use is
\begin{equation}
F=\frac{{\rm Tr}(UU^\dagger)+|{\rm Tr}(U^\dagger_{\rm target}U)|^2}{d(d+1)},
\end{equation} 
where the desired gate operation, $U_{\rm target}$, is in 2D logical space, the operation $U$ is the rotation operator $R$ projected onto 2D logical space, and $d=2$.

\end{widetext}

\end{document}